\documentclass[runningheads,a4paper]{llncs}
\usepackage{amsmath, epsfig, subfigure, url}

\begin{document}
\mainmatter

\title{Towards Reliable Automatic Protein Structure Alignment}
\author{Xuefeng Cui\inst{1} \and Shuai Cheng Li\inst{2} \and Dongbo Bu\inst{3} \and\\
Ming Li\inst{1} \thanks{Email: mli@cs.uwaterloo.ca}}
\institute{University of Waterloo, Ontario, Canada\\
\and City University of Hong Kong, Hong Kong, China\\
\and Chinese Academy of Sciences, Beijing, China}
\titlerunning{Towards Reliable Automatic Protein Structure Alignment}
\authorrunning{Xuefeng Cui, Shuai Cheng Li, Dongbo Bu, Ming Li}
\toctitle{Towards Reliable Automatic Protein Structure Alignment}
\tocauthor{Xuefeng Cui, Shuai Cheng Li, Dongbo Bu, Ming Li}
\maketitle

\begin{abstract}

A variety of methods have been proposed for structure similarity calculation,
which are called structure alignment or superposition. One major shortcoming in
current structure alignment algorithms is in their inherent design, which is
based on local structure similarity. In this work, we propose a method to
incorporate global information in obtaining optimal alignments and
superpositions. Our method, when applied to optimizing the TM-score and the GDT
score, produces significantly better results than current state-of-the-art
protein structure alignment tools. Specifically, if the highest TM-score found
by TMalign is lower than \(0.6\) and the highest TM-score found by one of the
tested methods is higher than \(0.5\), there is a probability of \(42\%\) that
TMalign failed to find TM-scores higher than \(0.5\), while the same
probability is reduced to \(2\%\) if our method is used. This could
significantly improve the accuracy of fold detection if the cutoff TM-score of
\(0.5\) is used.

In addition, existing structure alignment algorithms focus on structure
similarity alone and simply ignore other important similarities, such as
sequence similarity. Our approach has the capacity to incorporate multiple
similarities into the scoring function. Results show that sequence similarity
aids in finding high quality protein structure alignments that are more
consistent with eye-examined alignments in HOMSTRAD. Even when structure
similarity itself fails to find alignments with any consistency with
eye-examined alignments, our method remains capable of finding alignments
highly similar to, or even identical to, eye-examined alignments.
\end{abstract}

\section{Introduction}

Proteins function in living organisms as enzymes, antibodies, sensors, and
transporters, among myriad other roles. The understanding of protein function
has great implications to the study of biological and medical sciences. It is
widely accepted that protein function is determined mainly by structure.
Protein structures are often aligned for their common substructures, to
discover functionally or evolutionarily meaningful structure units. A very
large amount of data is available for such studies; the number of known protein
structures (the Protein Data Bank) has exceeded 90,000~\cite{pdb00}. Research
in structure alignments has intensified recently to enable efficient searches
of such databases.

Protein structures are usually modeled as 3-dimensional coordinates of atoms.
Thus, the alignment of two protein structures can be modeled as an optimization
problem to minimize the distance between two protein structures after a
specific rotation and translation. One problem with such comparisons is that
the time complexity is typically high. As a result, current methods for the
problem are heuristic in
nature~\cite{Akutsu1996,Nick96,Capr02,Mat04,Gers96,Gibr96,Lanc01,Sing97,STRUCTAL,CE,SOIPPA2}.

For example, TMalign~\cite{tmalign05} creates an initial alignment through
sequence and secondary structure alignments and extracts an initial
\emph{rotation and translation} (ROTRAN) accordingly. Then, the ROTRAN is
improved iteratively until convergence. This approach suffers from possibly
dissatisfactory initial alignments and from a lack of optimality guarantees in
the final results. TMalign was improved by the fragment-based approach in
fr-TM-align~\cite{frtmalign08}, in which local structure alignments are
computed and represented by the fragment alignments. A dynamic programming
technique is then employed to optimize the score function. However, this method
only guarantees the quality of the local alignment rather than of the global
alignment. 

An alignment of two subsets of residues (or \(C_{\alpha}\) atoms) corresponds
to a ROTRAN. Unlike fr-TM-align, we also consider the situation in which the
small sets contain remote residues. In addition, to overcome the problem of
computational inefficiency, we choose to filter the ROTRANs by clustering
rather than by using an exhaustive method. 

Experimental results suggest that both local fragments and remote fragment
pairs show significant contribution to finding higher
TM-scores~\cite{tmscore04} and to finding higher GDT scores~\cite{gdt99}, as
stated in Sections \ref{sec:exptms} and \ref{sec:expgdt}, respectively.
Specifically, if the highest TM-score found by TMalign~\cite{tmalign05} is
lower than \(0.6\) and the highest TM-score found by one of the tested methods
is higher than \(0.5\), there is a probability of \(42\%\) that TMalign failed
to find TM-scores higher than \(0.5\), while the same probability is reduced to
\(2\%\) with our method. Our method is also capable of finding alignments with
significantly (up to \(0.21\)) higher TM-scores. This could significantly
improve the accuracy of fold detection if the cutoff TM-score of \(0.5\) is
used.

Another limitation of current protein structure alignment scoring functions,
the TM-score~\cite{tmscore04} and the LG-score~\cite{lgscore98}, is that only
protein structure similarity is taken into consideration, while other important
protein similarities, such as sequence similarity, are ignored. It has been
observed that many protein structure alignments, based only on protein
structure similarity are highly sensitive to conformational
changes~\cite{review08}. Recently, sequence similarity has been incorporated
into the scoring function~\cite{formatt12,deepalign13}. In this paper we
introduce a new scoring function incorporating a variety of protein
similarities.

In Section \ref{sec:exphomstrad}, we demonstrate that sequence similarity
enables discovery of high quality protein structure alignments that are more
consistent with eye-examined alignments. Even when structure similarity itself
fails to find alignments with any consistency with eye-examined alignments in
HOMSTRAD~\cite{homstrad98}, our method is nevertheless able to find alignments
highly similar to, or even identical to, the eye-examined alignments. When the
aligned protein structures contain a high percentage of helices,
TM-score~\cite{tmscore04} involving only structure similarity sometimes cannot
avoid shifting the HOMSTRAD alignment by a few residues. In our experiment,
such shifting tends to be avoided by our scoring function, which involves both
structure and sequence similarities.

\section{Method}

Our protein structure alignment search method can be divided into two parts:
the search algorithm and the scoring function. In Section \ref{sec:search}, we
describe our search algorithm, which samples and selects near optimal
alignments reliably and efficiently. In Section \ref{sec:score}, we describe
our scoring function for evaluating the quality of an alignment accurately.

\subsection{Protein Structure Alignment Search Algorithm}
\label{sec:search}

Given a protein structure alignment scoring function, finding the optimal
alignment involves finding the optimal ROTRAN that maximizes the alignment
score. Assume that there exists a near optimal ROTRAN that minimizes the RMSD
of two small sets of \(C_{\alpha}\) atoms. We find the near optimal structure
alignment by sampling ROTRANs in four steps: (1) ROTRANs are initially sampled
from local fragment alignments and from remote fragment pair alignments; (2)
noise ROTRANs are filtered out by clustering; (3) one representative alignment
for each ROTRAN cluster is selected based on alignment scores; (4) the selected
alignments are refined by random ROTRAN sampling. Steps one through four are
discussed in this section and our scoring function is discussed in Section
\ref{sec:score}.

First, an initial set of ROTRANs must be sampled. Here, the primary concern is
to have several good candidates, instead of to have a high signal-to-noise
ratio, which is addressed in the next step. Finding good candidates is done by
calculating the optimal ROTRAN that minimizes RMSD between one or two fragments
from each protein structure. In case of a single fragment from each protein
structure, we call it local fragment. In case of two fragments from each
protein structure, we call them remote fragment pair. Here, we require the pair
of remote fragments to be of the same size and to be at least three residues
away from each other to avoid modeling information redundant to the local
fragments. In practice, a significantly large number of ROTRANs with the lowest
RMSDs are kept for the next step, and the actual number of ROTRANs is selected
empirically as stated in Section \ref{sec:exptms}.

Since the initial set of ROTRANs may contain a great deal of noise, we try to
filter out most of the noise with a star-like k-median clustering algorithm in
the second step. Assuming that we know the maximum distance \(\epsilon\)
between the median of a cluster and any member of the same cluster, an
approximate clustering is applied using a neighbor graph: each vertex
represents a rotation matrix, and two vertices are connected if and only if the
distance between them is at most \(\epsilon\). For each iteration, the vertex
with the highest degree and its neighbors are grouped into a cluster, and are
removed from the neighbor graph. The iteration repeats until either there are
no vertices of degree higher than one or until the maximum number of clusters
is reached. The unclustered ROTRANs are treated as noise. Similar approximate
clustering algorithms have been used~\cite{cluster04} and
studied~\cite{cluster09}.

To complete the clustering algorithm, we need a distance function
between ROTRANs. The Riemannian distance is a widely used distance metric
measuring the length of the shortest geodesic curve between two rotation
matrices~\cite{rotation02}. Since the transition vector can be calculated by
the rotation matrix and the weight centers of the aligned residues, we use
Riemannian distances between rotation matrices to avoid using redundant
information when clustering ROTRANs.

For each cluster, we find the representative alignment defined by the ROTRAN
that yields the highest alignment score within the cluster. The alignment score
is defined in Section \ref{sec:score}, and is calculated by the
Needleman-Wunsch dynamic programming algorithm~\cite{dp70}. Since dynamic
programming is computationally expensive, the number of clusters in the
previous step must be carefully determined to avoid wasting computation on
clusters of noise. After all alignment scores have been calculated, the top
scored alignments are selected for the refinement step.

Finally, we refine the selected representative alignments by random ROTRAN
sampling. Specifically, for each alignment to be refined, six aligned residue
pairs are randomly selected from the alignment, the ROTRAN that minimizes RMSD
of the aligned residue pairs is calculated, the alignment score of the
alignment defined by the sampled ROTRAN is also calculated, and the previous
steps are repeated until there are no improvements after \(l_1 l_2\)
iterations, where \(l_1\) and \(l_2\) are the number of residues of the two
aligned protein structures.

\begin{figure}
  \centering
  \subfigure[ROTRANs initially sampled]{
    \includegraphics[width=55mm]{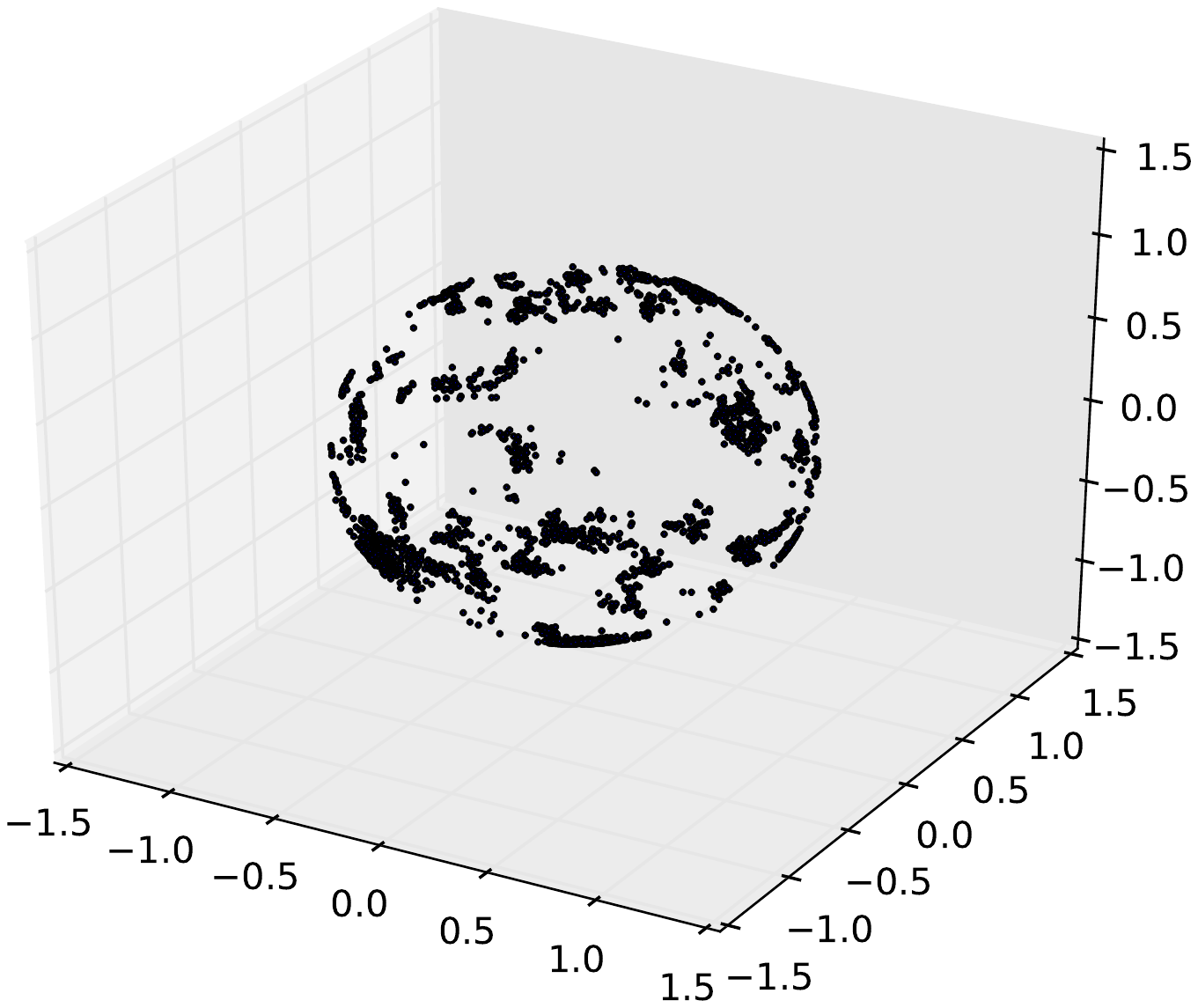}
    \label{fig:rotranall}
  }
  \subfigure[ROTRANs of the four largest clusters]{
    \includegraphics[width=55mm]{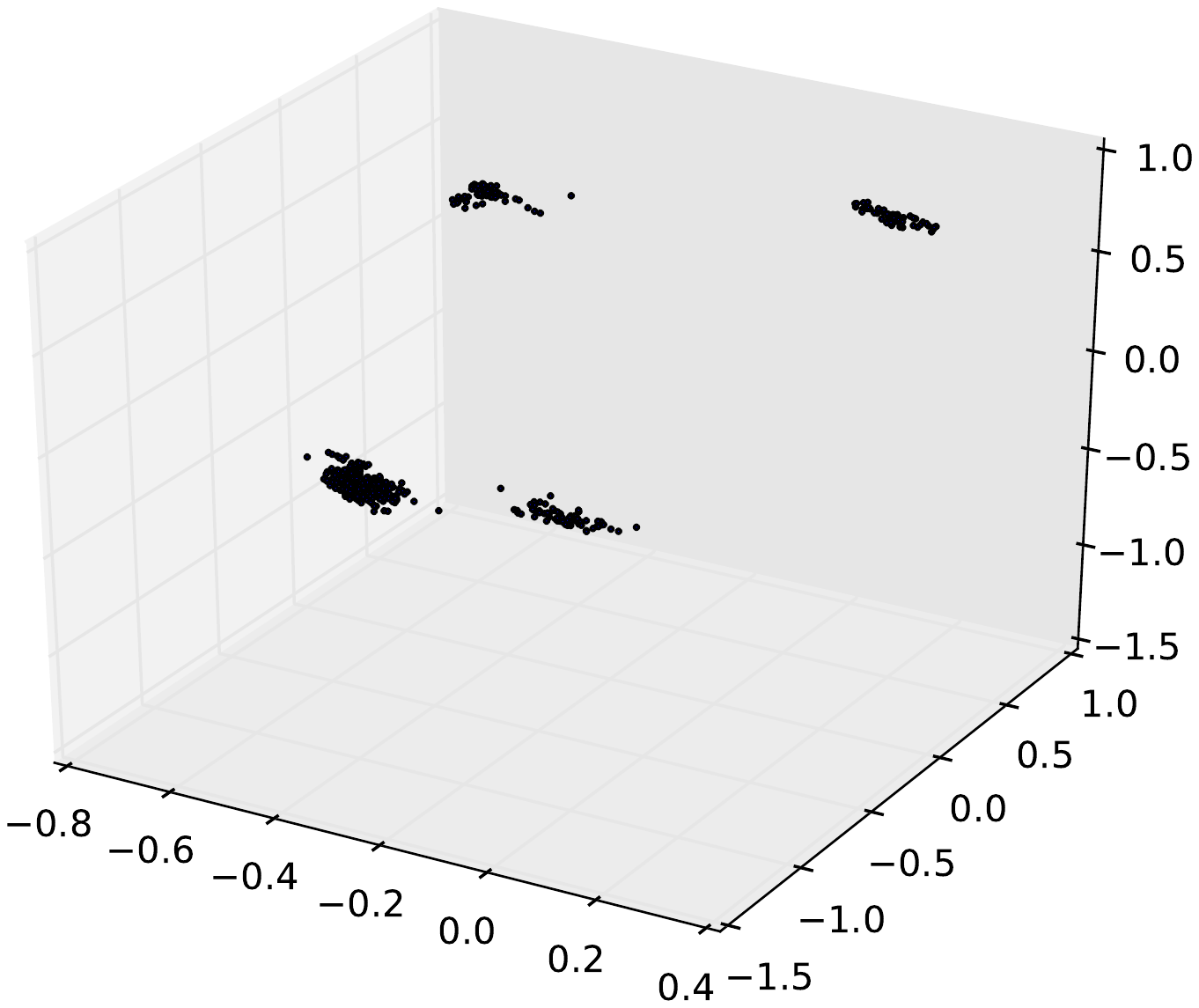}
    \label{fig:rotrancluster}
  }
  \caption{ROTRANs before and after clustering when aligning SCOP domains
  d3k2aa\_ and d2cufa1: each ROTRAN is represented by a coordinate that is
  calculated by applying the rotation matrix of the ROTRAN on coordinate \((1,
  0, 0)\).}
  \label{fig:rotran}
\end{figure}

The example shown in Figure \ref{fig:rotran} demonstrates the efficiency of our
protein structure alignment search algorithm, when aligning SCOP domains
d3k2aa\_ and d2cufa1~\cite{scop95}. In the figure, each coordinate represents a
ROTRAN because the coordinate is calculated by applying the rotation matrix of
the ROTRAN on the coordinate \((1, 0, 0)\). By looking at the initially sampled
ROTRANs shown in Figure \ref{fig:rotranall}, we can see that the ROTRANs have a
non-uniform distribution, and the ROTRANs with a small number of neighbors are
potential noise candidates. After clustering, the four largest clusters include
\(19\%\) of the initially sampled ROTRANs, as shown in Figure
\ref{fig:rotrancluster}. Note that the optimal ROTRAN that maximizes the
alignment score is located in the largest cluster, which includes \(13\%\) of
the initially sampled ROTRANs. Therefore, our search algorithm is highly
efficient because the alignment score calculation (by the computationally
expensive dynamic programming algorithm) for noise ROTRANs is mainly
eliminated. It is also possible to trade accuracy for speed by reducing the
number of sampled ROTRANs and reducing the number of clusters.

Our search algorithm is both efficient and reliable. Since similar protein
structures tend to have many local fragments, or remote fragment pairs with
small RMSDs, and similar rotation matrices, these rotation matrices tend to
form to a large cluster in our method. Since the rotation matrix space is
limited and we assume that the maximum distance between two rotation matrices
within a cluster is a constant, the maximum number of clusters within the
rotation matrix space is limited. This implies that the number of ROTRANs
required to accurately identify large clusters is also limited. Therefore, it
is only necessary to sample a limited number of ROTRANs, which is sufficient to
identify the large cluster containing near optimal ROTRANs.

\subsection{Protein Structure Alignment Scoring Function}
\label{sec:score}

TM-score~\cite{tmscore04}, based on LG-score~\cite{lgscore98}, is one of the
most successful protein structure alignment scoring functions. However, one
limitation of TM-score and LG-score is that they use only protein structure
similarity while they ignore other protein similarities, such as the sequence
similarity. It has been observed that many protein structure alignments, based
only on protein structure similarity, are highly sensitive to conformational
changes~\cite{review08}. This suggests the incorporation of other protein
similarities, such as the sequence similarity, in the protein structure
alignment scoring function. Here, we introduce a new scoring function
incorporating variety kinds of protein similarity as follows:
\[S = \displaystyle \dfrac{1}{L_r} \sum_{i \le l} \dfrac{1}{1 + f_a(D_1(i), D_2(i), ..., D_n(i))},\]
where \(L_r\) is the reference protein size; \(l\) is the number of aligned
residue pairs of the alignment; \(f_a\) is the weighted averaging function
(e.g. arithmetic, geometric or harmonic average); \(D_k(i)\) is the normalized
distance of the \(i\)-th aligned residue pair using the \(k\)-th distance
function; and \(n\) is the number of distance functions incorporated. If there
is \(n = 1\) and \(D_1(i) = (d_i/d_0)^2\), where \(d_i\) is the distance
between the \(C_{\alpha}\) atoms of the \(i\)-th aligned residue pair and
\(d_0\) is a normalization factor, our scoring function is identical to the
LG-score~\cite{lgscore98}. If there is also \(d_0 = 1.24 (L_r - 15)^{1/3} -
1.8\), our scoring function is identical to the TM-score~\cite{tmscore04}.
Thus, LG-score and TM-score are two special cases of our scoring function.

As an initial study on our new scoring function, we focus on the geometric
average of the normalized \(C_{\alpha}\) distance \(D_1(i)\) and the normalized
amino acid distance \(D_2(i)\) as follows:
\[S = \displaystyle \dfrac{1}{L_r} \sum_{i \le l} \dfrac{1}{1 + \sqrt[1+w]{D_1(i) D_2^w(i)}},\]
where \(w\) is a weighting factor. As with TM-score~\cite{tmscore04}, we define
the normalized \(C_{\alpha}\) distance as
\[D_1(i) = (\dfrac{d_i}{d_0})^2,\]
where \(d_0 = 1.24 (L_r - 15)^{1/3} - 1.8\). Based on the popular BLOSUM62
matrix~\cite{blosum92,blosum04}, we define the normalized amino acid distance
as
\[D_2(i) = \displaystyle 2^{-M(P_i, Q_i)}
         = \displaystyle 2^{-\lambda \log \frac{P(P_i, Q_i)}{P(P_i) P(Q_i)}}
         = \displaystyle (\frac{P(P_i) P(Q_i)}{P(P_i, Q_i)})^{\lambda},\]
where \(M\) is the BLOSUM62 matrix, \((P_i, Q_i)\) is the \(i\)-th aligned
residue pair, \(\lambda\) is a scaling factor, \(P(P_i, Q_i)\) is the
probability of amino acid \(P_i\) aligning to amino acid \(Q_i\), and
\(P(P_i)\) and \(P(Q_i)\) are the probabilities of amino acid \(P_i\) and amino
acid \(Q_i\), respectively. Instead of using the default scaling factor
\(\lambda\), it is treated here as a parameter to control the rate of mutation.

An appealing property shared between TM-score~\cite{tmscore10} and our scoring
function is that the in-favored protein structure alignments tend to have
scores higher than \(0.5\). If the \(C_{\alpha}\) distance between the \(i\)-th
aligned residue pair is in-favored, there is \(d_i < d_0\) and thus \(D_1(i) <
1\). If the amino acid distance between the \(i\)-th aligned residue pair is
in-favored, there is \(P(P_i, Q_i) > P(P_i) P(Q_i)\) and thus \(D_2(i) < 1\).
Then, for the \(i\)-th aligned residue pair, there is \(D_1(i) D_2(i) < 1\) and
thus \(1 / (1 + \sqrt[1+w]{D_1(i) D_2^w(i)}) > 0.5\). Therefore, if many
in-favored aligned residue pairs occur in the alignment, our protein structure
alignment score tends to be higher than \(0.5\).


\section{Result}

We included three experiments to demonstrate that the protein structure
alignments found by using our method are not only higher scored but are also
more consistent with those alignments examed visually by human-beings. In
Section \ref{sec:exptms}, we compared our search algorithm to current
state-of-the-art search algorithms, TMalign~\cite{tmalign05} and
fr-TM-align~\cite{frtmalign08}, to demonstrate that our method tends to find
alignments with higher TM-scores~\cite{tmscore04}. In Section \ref{sec:expgdt},
we compared our search algorithm to SPalign~\cite{spalign12} to demonstrate
that our method tends to find alignments with higher GDT scores~\cite{gdt99}.
In Section \ref{sec:exphomstrad}, we compared our scoring function to
TM-score~\cite{tmscore04} to demonstrate that our method tends to find
alignments more consistent with the eye-examined alignments in
HOMSTRAD~\cite{homstrad98}.

\subsection{Search Algorithm Evaluation on TM-score}
\label{sec:exptms}

To demonstrate reliability, we repeated the alignment experiment for the
\(200\) non-homologous protein structures, which have sizes of between \(46\)
and \(1058\), have a sequence identity cutoff of \(30\%\), and are used by
TM-align~\cite{tmalign05}. We compared our results with that of current
methods, TM-align and fr-TM-align~\cite{frtmalign08}. Here, we used
TM-score~\cite{tmscore04} normalized by the smaller protein size as the scoring
function. Since fr-TM-align does not support normalization by the smaller
protein size, TM-score normalized by the smaller protein size is calculated
based on the rotation matrix returned by fr-TM-align. Since biologists tend to
be more interested in similar protein structures within the same protein fold,
and the TM-score of \(0.5\) is a good approximate threshold for protein fold
detection~\cite{tmscore10}, only the \(350\) protein structure alignments with
TM-scores higher than \(0.5\) (found by at least one of the tested methods) are
included in this analysis.

For the experiment settings in the algorithm described in Section
\ref{sec:search}, we used local fragments of size \(12\), and remote fragment
pairs of size \(3\). Such experiment settings are called L12R3align. To study
the contributions of using local fragments and using remote fragment pairs, we
simplified our method to two variants: L12align, that used only local fragments
of size \(12\), and R3align, that used only remote fragment pairs of size
\(3\). For consistency, we selected \(1,536\) local fragments of size \(12\)
and \(1,536\) remote fragment pairs of size three in the sampling step, used
\(\epsilon = 10^{\circ}\) in the clustering step, stopped clustering when
\(288\) clusters were found, and selected eight clusters in the refinement step
in all experiments for this section. With L12R3align, the elapsed time required
to finish this experiment was approximately \(4.5\) hours on a computer with
dual Intel Xeon X5660 2.8GHz CPUs and dual Nvidia GeForce GTX 670 GPUs. Thus,
each pairwise alignment took approximately \(0.8\) seconds on average.

\begin{figure}
  \centering
  \subfigure[Cluster rank containing the optimal ROTRAN]{
    \includegraphics[width=55mm]{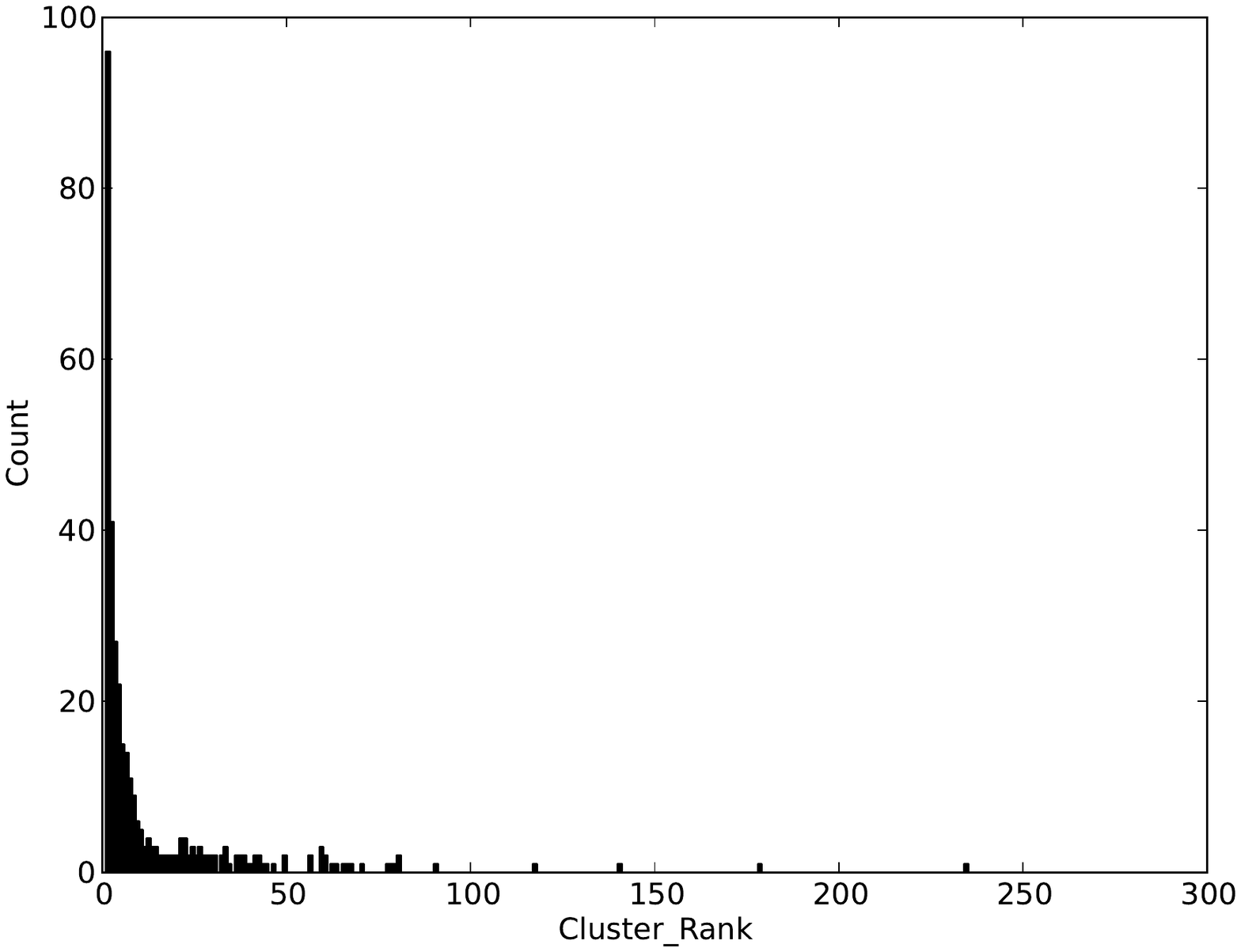}
    \label{fig:cluster}
  }
  \subfigure[TM-score before and after refinement]{
    \includegraphics[width=55mm]{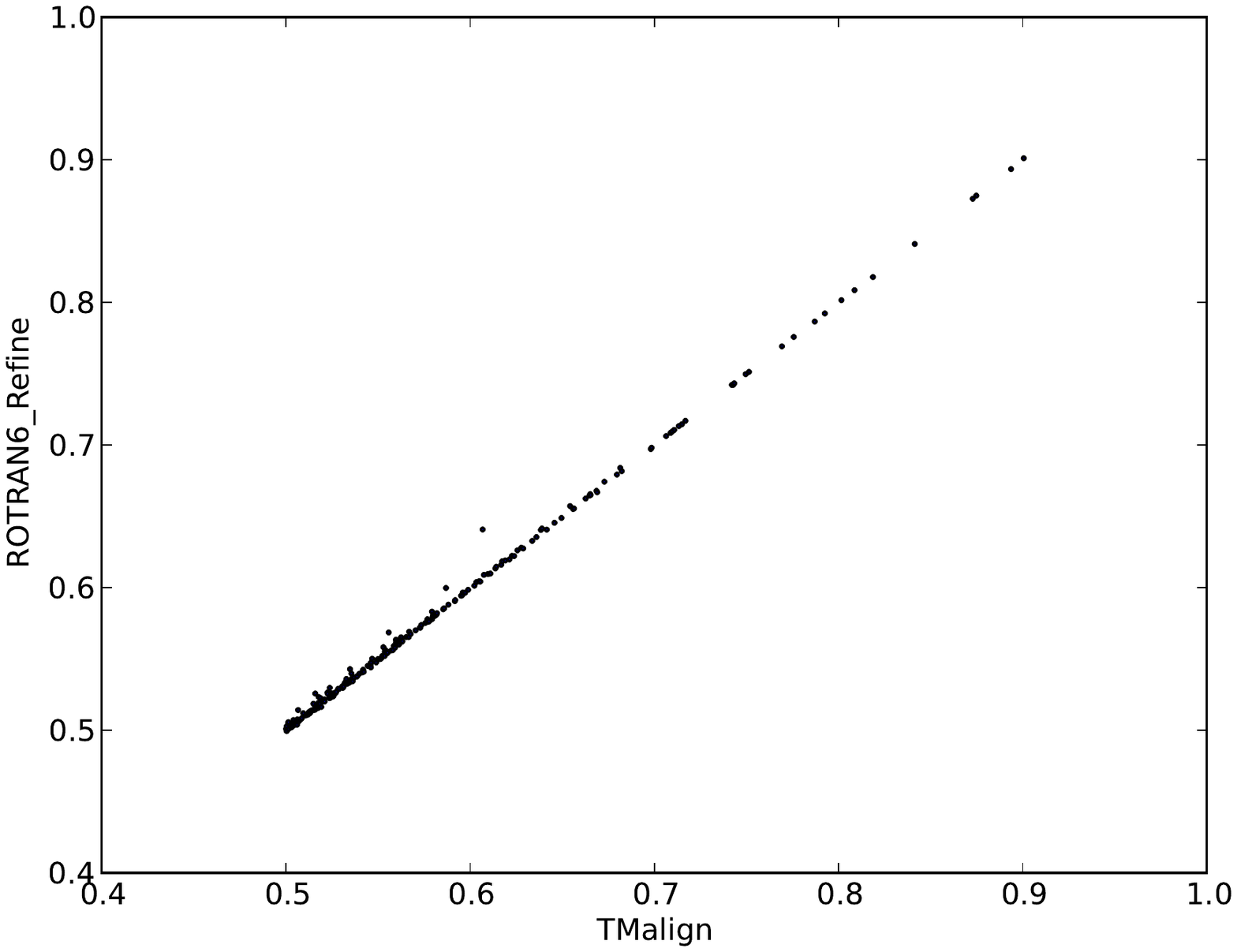}
    \label{fig:refine}
  }
  \subfigure[TMalign v.s. L12align]{
    \includegraphics[width=55mm]{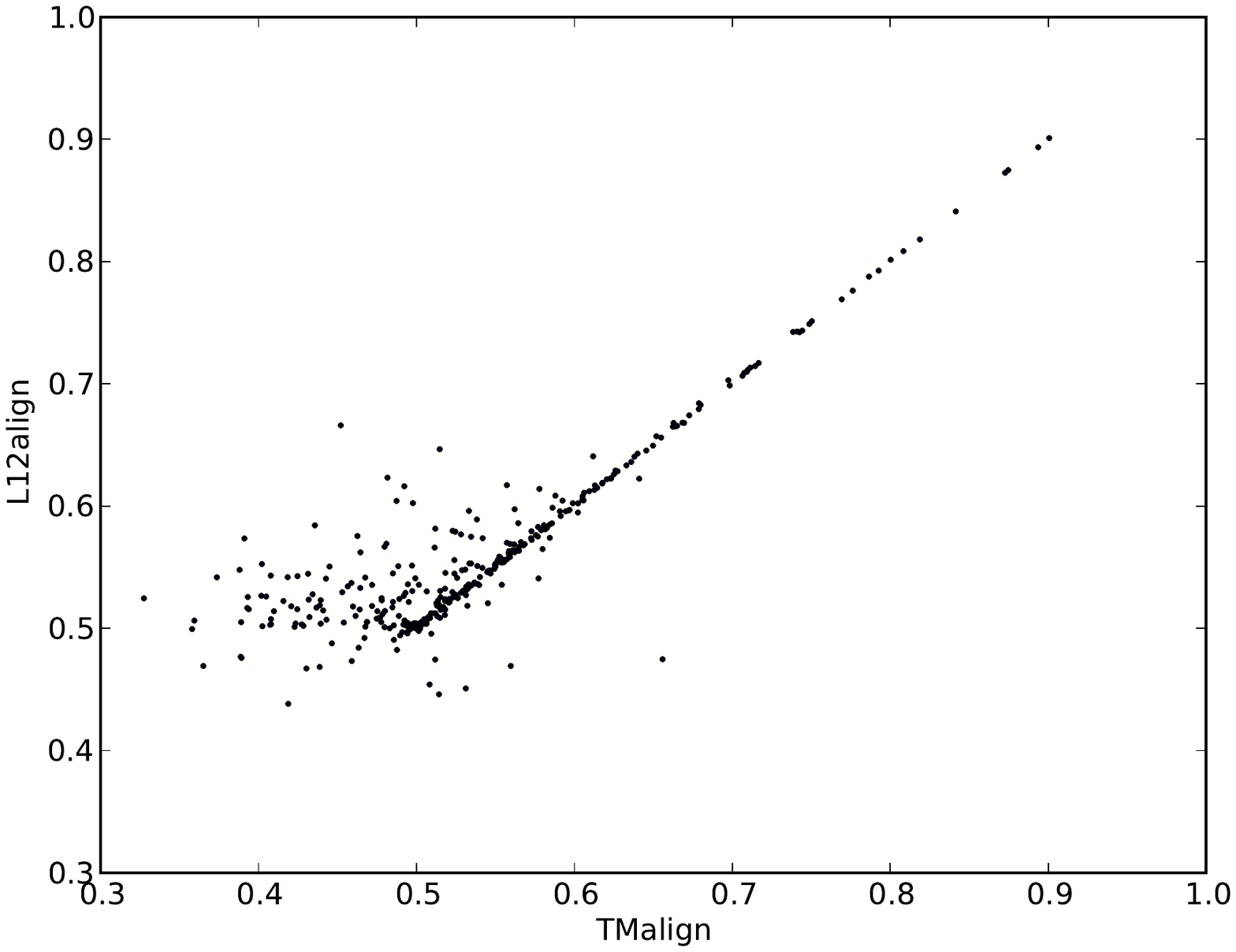}
    \label{fig:tmlc}
  }
  \subfigure[TMalign v.s. R3align]{
    \includegraphics[width=55mm]{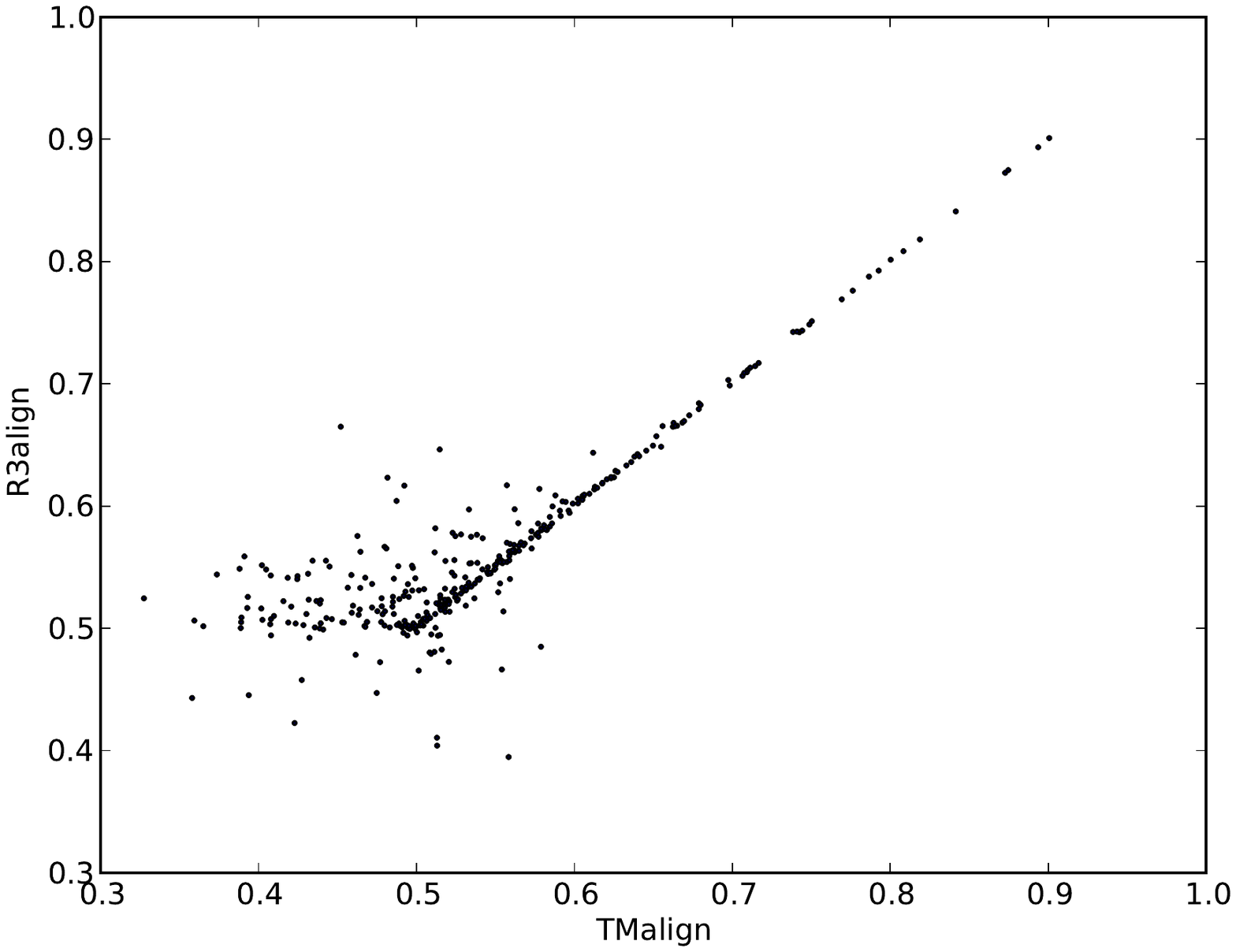}
    \label{fig:tmrm}
  }
  \subfigure[TMalign v.s. L12R3align]{
    \includegraphics[width=55mm]{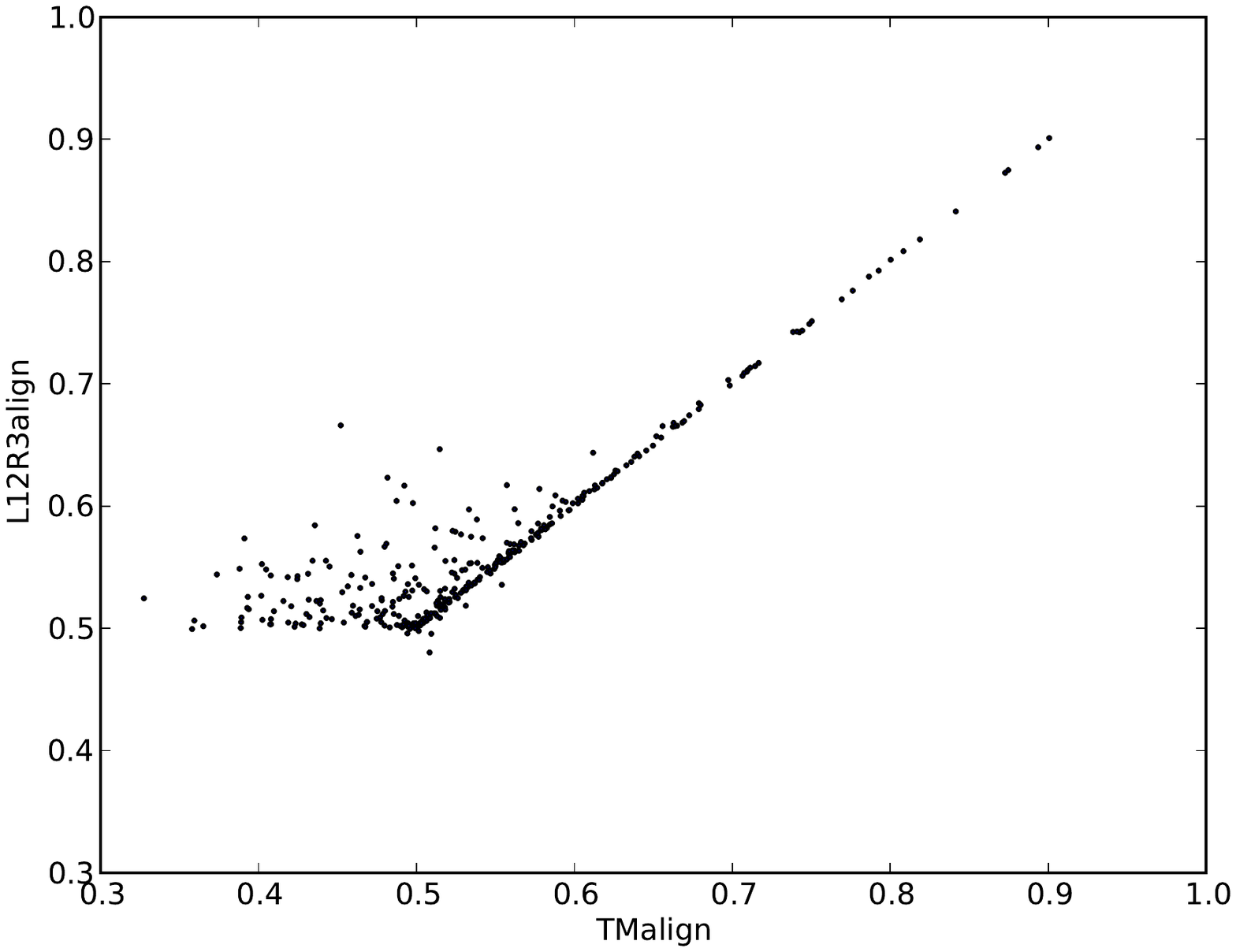}
    \label{fig:tmlcrm}
  }
  \subfigure[fr-TM-align v.s. L12R3align]{
    \includegraphics[width=55mm]{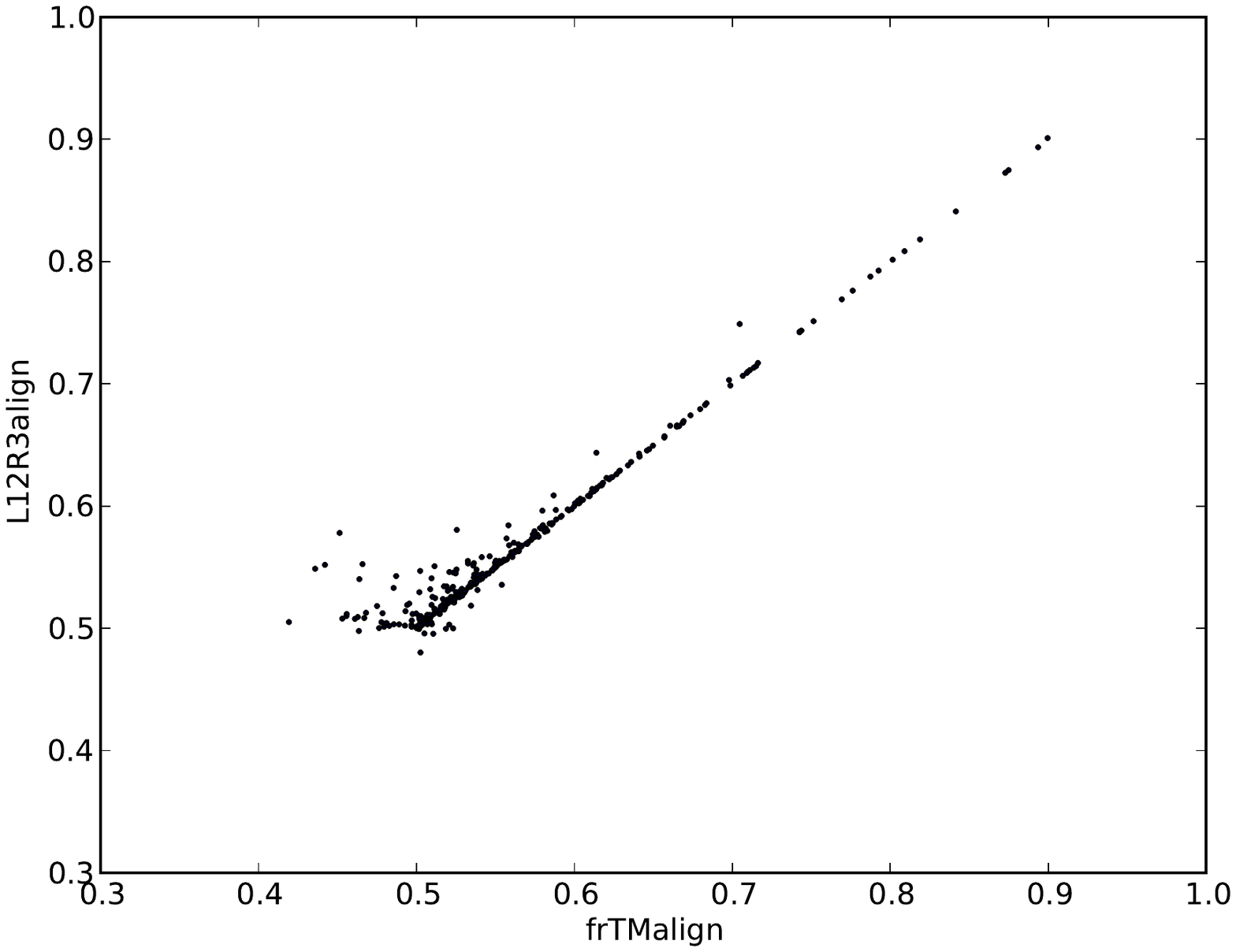}
    \label{fig:frlcrm}
  }
  \caption{Comparisons of the highest TM-scores found by TMalign and by using our method}
\end{figure}

First, we would like to evaluate the ROTRAN filtering step described in Section
\ref{sec:search}. Figure \ref{fig:cluster} shows the cluster rank that contains
the optimal ROTRAN with the highest TM-score~\cite{tmscore04}. Here, we focus
on the results of using local fragments because the results of using remote
fragment pairs draws similar conclusions. Specifically, \(28\%\) of the optimal
ROTRANs are from the largest cluster and \(72\%\) of the optimal ROTRANs are
from the largest ten clusters. Moreover, only \(1\%\) of the optimal ROTRANs
are not from the largest \(100\) clusters. This demonstrates that the optimal
ROTRAN tends to have many similar ROTRANs that minimize the RMSD of local
fragment alignments, and that these ROTRANs tend to form a large cluster, which
can be identified easily by clustering the sampled ROTRANs.

Next, we will demonstrate that our refinement step using randomly selected
ROTRANs, as described in Section \ref{sec:search}, is able to consistently find
protein structure alignments with similar or higher TM-scores~\cite{tmscore04}.
Figure \ref{fig:refine} shows the TM-score before and after refining the
optimal alignment found by TMalign~\cite{tmalign05}. It can be seen that the
TM-scores are mostly similar, while our refinement occasionally improves the
TM-score by up to \(0.10\). Specifically, after refinement, all TM-scores are
at most \(0.0029\) lower, while \(3\%\) of the TM-scores are at least \(0.01\)
higher. Recall that the random ROTRANs used in the refinement step are
generated by finding the ROTRAN that minimizes the RMSD of size randomly
selected aligned residue pairs from the alignment. Thus, this result also
verifies our assumption that there exists a near optimal ROTRAN that minimizes
the RMSD of two small sets of \(C_{\alpha}\) atoms.

To support our choices of local fragment size and of remote fragment pair size,
the highest TM-scores found by L12align and R3align are compared to those found
by TMalign in Figures \ref{fig:tmlc} and \ref{fig:tmrm}, respectively. For
protein structure pairs that have TMalign TM-scores higher than \(0.6\), both
L12align and R3align can reliably find high quality alignments with similar
TM-scores. For the other protein structure pairs, both L12align and R3align
tend to improve TM-scores, although there may be some reductions of TM-scores.
This demonstrates that both L12align and R3align are capable of finding high
quality alignments that are comparable to or even better than those found by
TMalign. In fact, the local fragment size of \(12\) has also been used by
fr-TM-align~\cite{frtmalign08}.

The improvements of TM-scores found by L12R3align over those found by TMalign
are shown in Figure \ref{fig:tmlcrm}. We see that TM-scores found by L12R3align
are mainly higher than those found by TMalign for the \(284\) protein structure
pairs that have TMalign TM-scores lower than \(0.6\). Specifically, L12R3align
improves TM-scores by \(0.03\) on average and by \(0.21\) in the best case.
Moreover, \(14\%\) of the TM-scores are improved by at least \(0.1\), \(30\%\)
of the TM-scores are improved by at least \(0.05\), and only \(2\%\) of the
TM-scores are reduced by at most \(0.03\). Comparing to Figures \ref{fig:tmlc}
and \ref{fig:tmrm}, the number of TM-scores found by our method that are lower
than those found by TMalign is significantly reduced using both local fragments
and remote fragment pairs.

If the highest TM-score found by TMalign is lower than \(0.6\) and the highest
TM-score found by one of the tested methods is higher than \(0.5\), there is a
probability of \(42\%\) that TMalign failed to find TM-scores higher than
\(0.5\). In such cases, L12R3align tends to discover better protein structure
alignments with (possibly significantly) higher TM-scores, with a probability
of only \(2\%\) that L12R3align failed to find TM-scores higher than \(0.5\).
This could significantly improve fold detection results. Interestingly,
L12R3align tends to improve TM-scores more for \(\alpha\)-proteins, while
never reduces TM-scores for \(\beta\)-proteins.

In addition to comparison with TMalign, the TM-scores found by L12R3align are
also compared with those found by fr-TM-align~\cite{frtmalign08} as shown in
Figure \ref{fig:frlcrm}. Note that TM-scores found by L12R3align are also
mainly higher than those found by fr-TM-align for protein structure pairs that
have fr-TM-align TM-scores lower than \(0.6\). Specifically, L12R3align
improves TM-scores by up to \(0.13\), while it reduces TM-scores by at most
\(0.02\). Moreover, L12R3align finds \(28\) more TM-scores that are higher than
\(0.5\).


\subsection{Search Algorithm Evaluation on GDT Score}
\label{sec:expgdt}

In addition to TM-score~\cite{tmscore04}, GDT~\cite{gdt99} score is also one of
the most popular protein structure alignment scoring function~\cite{casp9}.
Thus, we repeated the experiment in Section \ref{sec:exptms}, but compared the
GDT scores found by our method to those found by SPalign~\cite{spalign12},
which is a new protein structure alignment tool that uses a search algorithm
similar to that of TMalign. SPalign aims to find one of the highest SP-score,
the highest TM-score, or the highest GDT score. If we included SPalign in the
previous experiment in Section \ref{sec:exptms}, it would perform slightly
better than TMalign on average. Thus, SPalign has a effective search algorithm
and it should be a candidate for finding the highest GDT score for
comparison. Again, only the \(339\) protein structure alignments with GDT
scores higher than \(0.5\), found by at least one of the tested methods, are
included in this analysis.

\begin{figure}
  \centering
  \subfigure[SPalign v.s. L12R3align]{
    \includegraphics[width=55mm]{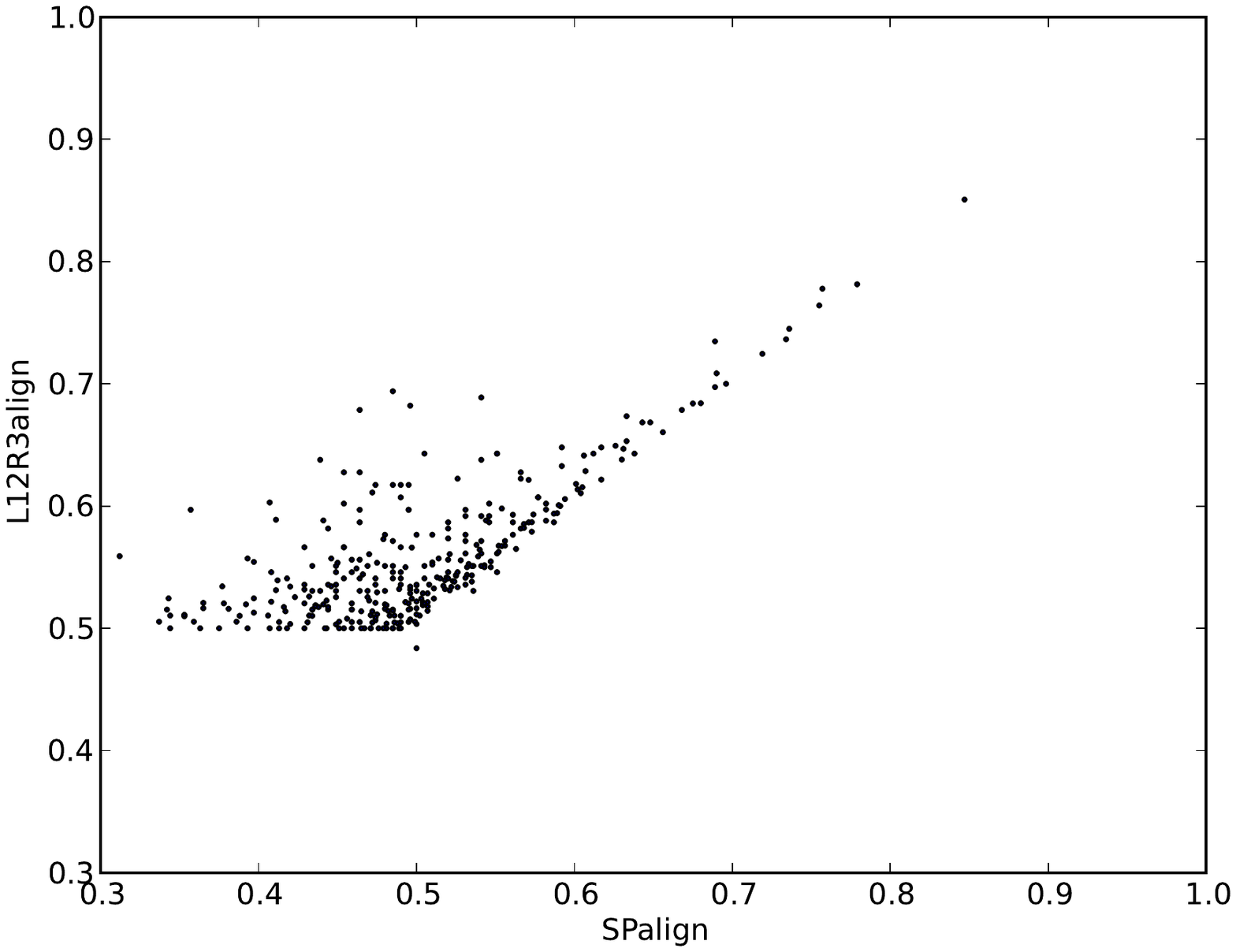}
    \label{fig:splcrm}
  }
  \subfigure[L12align v.s. R3align]{
    \includegraphics[width=55mm]{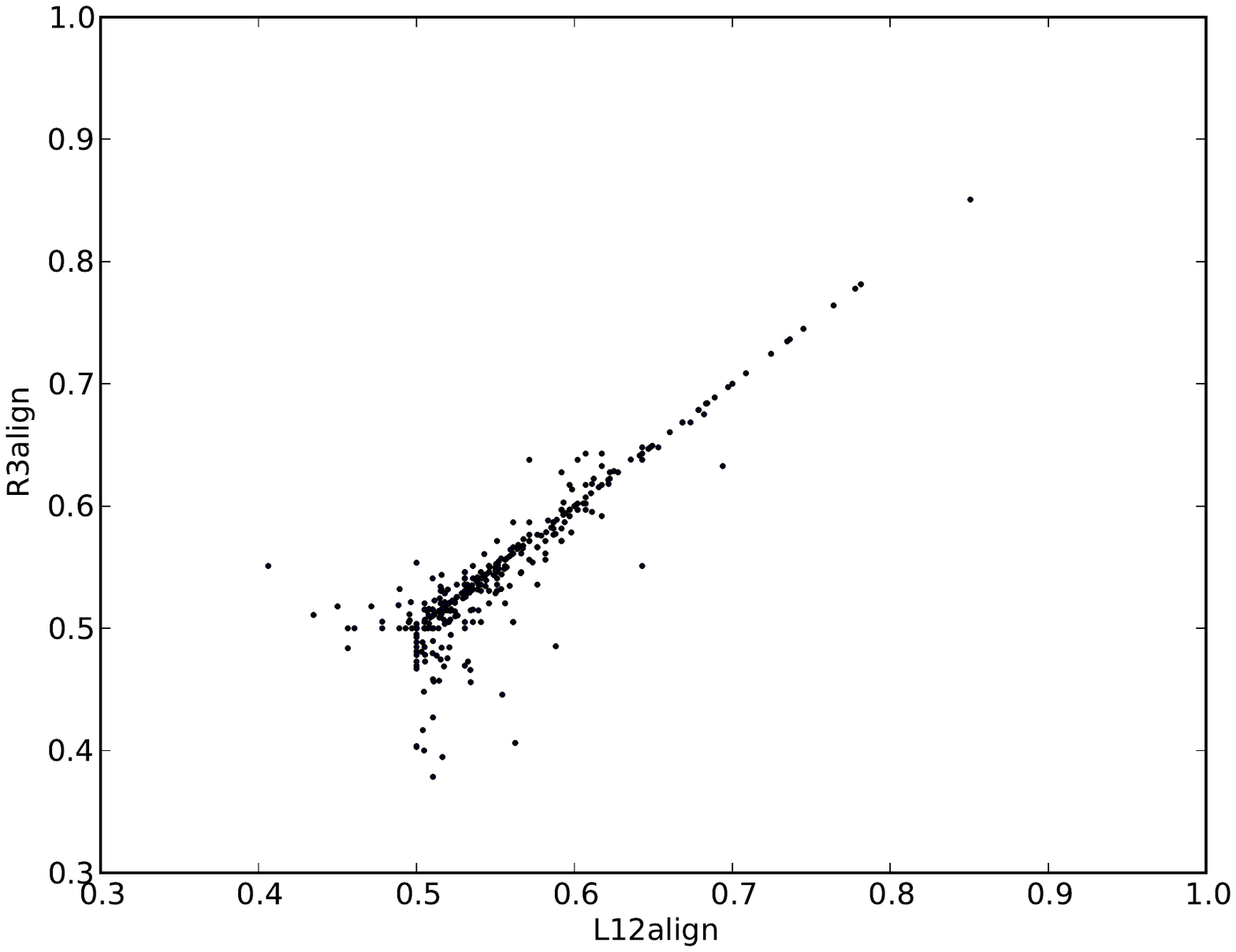}
    \label{fig:lcrm}
  }
  \caption{Comparisons of the highest GDT scores found by SPalign and by using our method}
\end{figure}

Comparing the GDT scores found by L12R3align and SPalign as shown in Figure
\ref{fig:splcrm}, we find that L12R3align consistently finds similar or higher
GDT scores than SPalign. Specifically, L12R3align improves GDT scores by
\(0.06\) on average and by \(0.25\) in the best case. It is seen that \(25\%\)
of the GDT scores are improved by at least \(0.09\) and that \(75\%\) of the
GDT scores are improved by at least \(0.02\). Moreover, SPalign finds \(145\)
alignments with GDT scores higher than \(0.5\), while L12R3align finds \(314\)
alignments with GDT scores higher than \(0.5\). Thus, \(169\) more alignments
with GDT scores higher than \(0.5\) are discovered, with an average GDT score
improvement of \(0.09\). These results again supports that our protein
structure alignment search algorithm can reliably find high quality alignments.

To further study the contributions of local fragments and remote fragment pairs
to the GDT score improvements of L12R3align over SPalign, the GDT scores found
by L12align and R3align are compared in Figure \ref{fig:lcrm}. It can be seen
that both L12align and R3align find similar GDT scores when one of the GDT
scores found by L12align and R3align is higher than \(0.65\). For the remaining
protein structure pairs, both L12align and R3align are capable of discovering
some better GDT scores than is the other method. Generally, \(47\%\) of the GDT
scores found by L12align are up to \(0.16\) higher and \(30\%\) of the GDT
scores found by R3align are up to \(0.14\) higher. Therefore, local fragments
have a greater contribution in finding the highest GDT scores, while remote
fragment pairs still have a significant contribution in finding the highest GDT
scores.


\subsection{Scoring Function Evaluation on Consistency with Eye-Examed Alignments}
\label{sec:exphomstrad}

In this experiment, we would like to show that our scoring function is capable
of finding protein structure alignments that are significantly more consistent
with alignments examed visually by human-beings. Thus, we used protein
structure alignments from the HOMSTRAD database~\cite{homstrad98} as a
benchmark and compared the alignment quality of our protein structure alignment
with that of TMalign~\cite{tmalign05}. Here, the quality of the alignment is
evaluated by the F-score, the harmonic mean of recall and precision, of aligned
residue pairs.

The HOMSTRAD database has been widely used in protein research, including
sequence-sequence alignment~\cite{probcons05}, sequence-structure
alignment~\cite{fugue01}, and structure-structure alignment~\cite{mustang06},
among others. The database contains structure alignments of \(3,454\)
homologous protein structures from \(1,032\) protein
families~\cite{homstrad98}. Since different sequences were read from alignment
files and from PDB structure files for some proteins, only \(9,429\) out of
\(9,535\) protein structure alignments from HOMSTRAD were included in this
experiment.

For our experiment settings, we chose \(\lambda = 0.25\) and \(w = 1.9\),
empirically. Unlike previous experiment settings, we used local fragments of
size \(9\) and remote fragment pairs of size \(3\). Such experiment settings
are balanced between the accuracy and the speed of our protein structure
alignment algorithm because only a minor improvement on accuracy is gained by
increasing the sizes, while slowing down the running time. The local fragment
size of \(9\) was previously shown to be the optimal balance between the
complexity of the model and the amount of data required to train the
model~\cite{rosetta04,xin12}. Other experiment settings remained the same as in
the previous experiment.

\begin{figure}
  \centering
  \subfigure[F-score difference between L9R3align and TMalign]{
    \includegraphics[width=55mm]{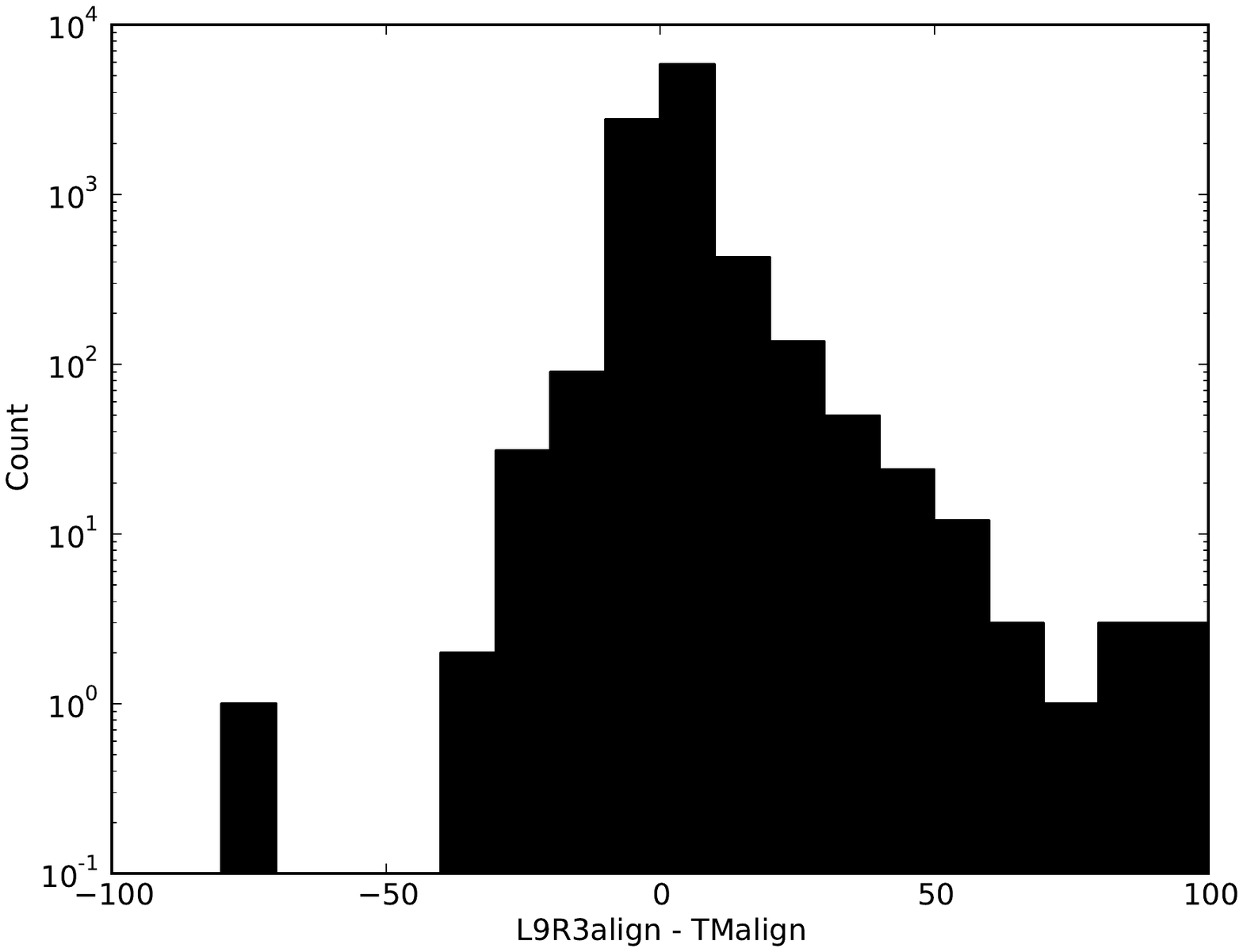}
    \label{fig:fdiff}
  }
  \subfigure[TMalign F-score v.s. L9R3align F-score]{
    \includegraphics[width=55mm]{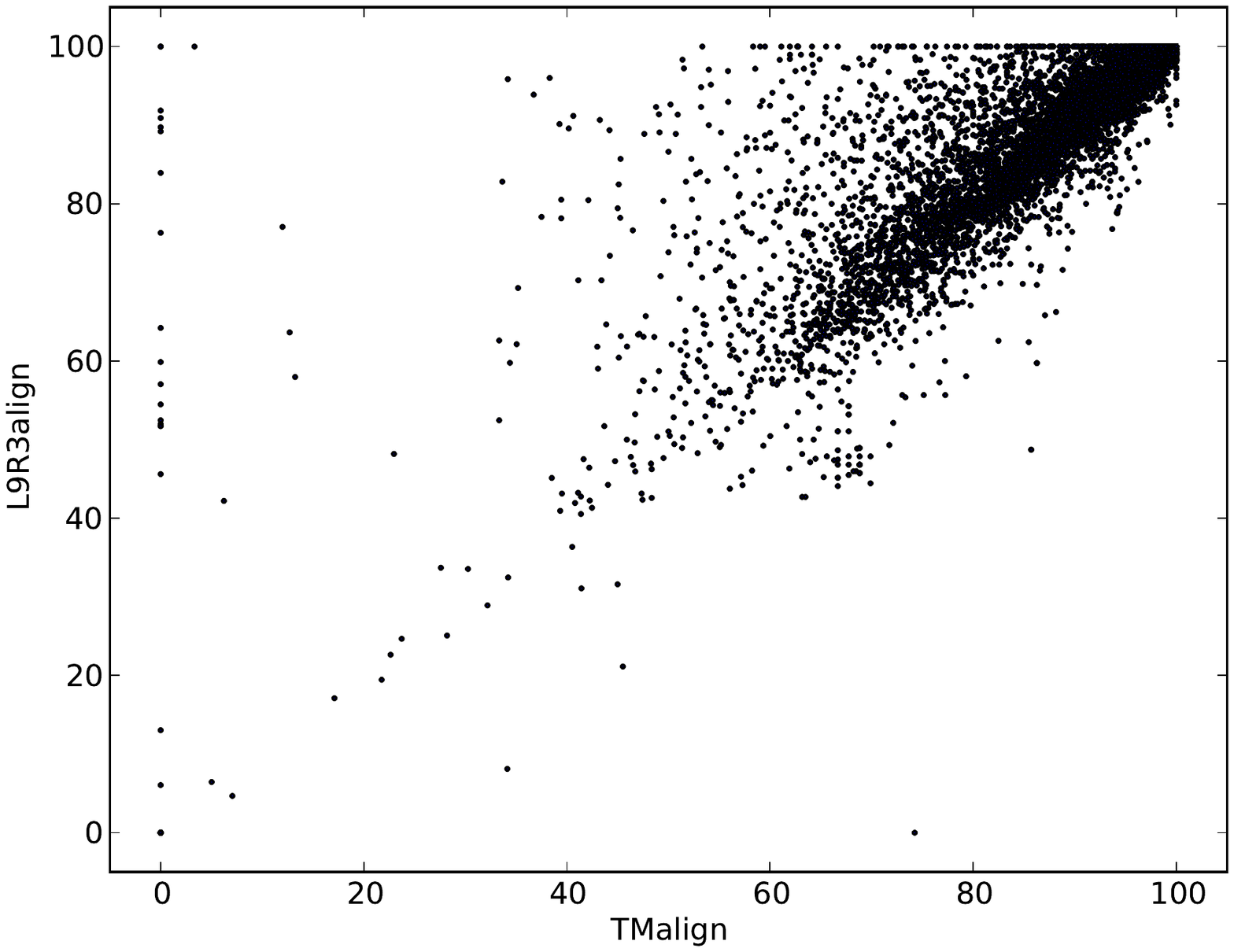}
    \label{fig:fscore}
  }
  \caption{Comparisons of the F-scores of the aligned residue pairs found by L9R3align and TMalign}
\end{figure}

The F-score differences between L9R3align and TMalign are shown in Figure
\ref{fig:fdiff}. Using L9R3align, \(47\%\) of the F-scores are improved, and
the average F-score is improved from \(88\%\) to \(90\%\) compared to using
TMalign. Moreover, there are \(663\) L9R3align F-scores that are at least
\(10\%\) higher and there are \(1,342\) L9R3align F-scores that are at least
\(5\%\) higher than the TMalign F-scores. For comparison, \(31\%\) of the
TMalign F-scores are higher, and only \(124\) TMalign F-scores are at least
\(10\%\) higher. In total, TMalign finds \(5,560\) protein structure alignments
with F-scores higher than \(90\%\), while L9R3align finds \(6,114\) such
alignments. Therefore, the protein structure alignments found by L9R3align are
\(10\%\) more likely to be highly consistent (with F-score higher than
\(90\%\)) with eye-examined alignments, and tend to have similar or higher
F-scores compared to the protein structure alignments found by TMalign.

Among the \(34\) pairs of protein structures that have TMalign F-scores equal
to zero as shown in Figure \ref{fig:fscore}, the L9R3align F-scores reach
\(36\%\) on average. Specifically, two L9R3align F-scores equal to \(100\%\)
and \(19\) L9R3align F-scores are higher than \(50\%\). For the two cases that
L9R3align F-scores are equal to \(100\%\), the aligned protein structures
contain a high percentage of helices, and TMalign shifts the HOMSTRAD alignment
by a few residues, which has also been previously observed~\cite{review12}.
Such shifting is difficult to avoid by evaluating only structure similarities.
However, the shifting is avoided by our scoring function, involving both
structure and sequence similarities, in this experiment. Therefore, sequence
similarity does aids in finding high quality protein structure alignments that
are highly consistent with eye-examined alignments, even if structure
similarity itself fails to do so.

There is also one pair of protein structures in Figure \ref{fig:fscore} that
the L9R3align F-score equals to zero, while the TMalign F-score equals to
\(74\%\). Here, the HOMSTRAD alignment can be represented by protein ``AB-''
aligning to protein ``-CD'', where each character represents a protein fragment
and ``-'' represents a gap region. One possible reason for this is that the
weight parameters of our scoring function are not yet optimized to completely
break the dependency between the alignment score and the protein size. We have
observed that such cases can be eliminated by using different weight
parameters, and this problem will be addressed in our future work.


\section{Discussion and Conclusion}

Therefore, our protein structure alignment method is not only reliable in
finding the optimal alignment with the highest alignment score, but is also
capable of discovering new alignments missed by current stat-of-art alignment
search algorithms and scoring functions. Our result verifies our assumption
that there exists a near optimal ROTRAN that minimizes the RMSD of two small
sets of \(C_{\alpha}\) atoms. Our result also verifies that although structure
similarity may be efficient in many cases, sequence similarity helps to find
better protein structure alignments that are (possibly significantly) more
consistent with eye-examined alignments. This is the result of incorporating
both local fragments and remote fragment pairs in the alignment search
algorithm, and of incorporating both structure similarity and sequence
similarity in the scoring function.

Our protein structure alignment algorithm is still subject to improvement and
application. Our scoring function remains capable of modeling more types of
protein similarities, such as the \((\phi, \psi)\) dihedral angle distance and
the secondary structure distance. Unknown protein domain length problems when
aligning multi-domain proteins should also be addressed in the future as
proposed by SPalign~\cite{spalign12}. It should be interesting to allow
flexible ROTRANs within the same cluster to find flexible structure alignments
as seen in FATCAT~\cite{fatcat03} and to find flexible multi-structure
alignments as seen in Matt~\cite{matt08}. Moreover, the alignment quality can
be further studied by evaluating CASP protein structure
prediction~\cite{casp9}, by checking self-consistency~\cite{review12}, and by
simulating the SCOP fold detection~\cite{scop95}. All these aid in fully
automating protein structure alignment process as good as or even better than
human experts in the short future.

{\small
  \subsubsection*{Acknowledgments:} This work was supported by the Startup
  Grant at City University of Hong Kong [7002731], the National Basic Research
  Program of China [2012CB316500], an NSERC Grant [OGP0046506], the Canada
  Research Chair program, an NSERC Collaborative Grant, OCRiT, the Premier's
  Discovery Award, the Killam Prize and SHARCNET.

  \bibliographystyle{splncs}
s  \bibliography{main}
}

\end{document}